\documentclass[preprint2,times,tighten]{aastex6}

\usepackage{color}
\usepackage{xcolor}
\usepackage{enumitem}
\usepackage{hyperref}
\usepackage{verbatim}
\usepackage{amsmath,amstext,mathrsfs}
\usepackage[all]{hypcap} 
\usepackage{afterpage}   
\usepackage{marginnote}
\usepackage{makecell}
\usepackage{multirow}

\begin{document}
\title{Tracing Multi-Scale Magnetic Field Structure Using Multiple Chemical Tracers in Giant Molecular Clouds}
\author{Yue Hu\altaffilmark{1,2}, Ka Ho Yuen\altaffilmark{1}, A. Lazarian\altaffilmark{1}, Laura M.Fissel\altaffilmark{3}, P. A. Jones\altaffilmark{4}, M. R. Cunningham\altaffilmark{4}}
\email{yue.hu@wisc.edu}
%\email{lazarian@astro.wisc.edu}
\altaffiltext{1}{Department of Astronomy, University of Wisconsin-Madison, Madison, WI, USA}
\altaffiltext{2}{Department of Physics, University of Wisconsin-Madison, Madison, WI, USA}
\altaffiltext{3}{National Radio Astronomy Observatory, Charlottesville, VA, USA}
\altaffiltext{4}{School of Physics, University of New South Wales, Sydney, NSW 2052, Australia}

\begin{abstract}
Probing magnetic fields in Giant Molecular Clouds is often challenging. Fortunately, recently simulations show that analysis of velocity gradients (the Velocity Gradient Technique) can be used to map out the magnetic field morphology of different physical layers within molecular clouds when applied CO isotopologues with different optical depths. Here, we test the effectiveness of the Velocity Gradient Technique in reconstructing the magnetic field structure of the molecular cloud Vela C,  employing seven chemical tracers that have different optical depths, i.e. $^{12}$CO, $^{13}$CO, C$^{18}$O, CS, HNC, HCO$^{+}$, and HCN. Our results show good correspondence between the magnetic field morphology inferred from velocity gradients using these different molecular tracers and the magnetic field morphology inferred from BLASTPol polarization observations. We also explore the possibility of using a combination of velocity gradients for multiple chemical tracers to explain the structure of the magnetic field in molecular clouds. We search for signatures of gravitational collapse in the alignment of the velocity gradients and magnetic field and conclude that collapsing regions constitute a small fraction of the cloud.
\end{abstract}

\section{Introduction}
The magnetic field in the universe plays an essential role in multiple astrophysical processes, e.g., regulation of star formation \citep{2015ApJ...808...48B,2013ApJ...770..151C}, guiding material and thermodynamic transfer between different media, and propagation and acceleration of cosmic rays \citep{1949PhRv...75.1169F,2014ApJ...783...91C}. One of the most critical roles of the magnetic field plays is to modify ubiquitous interstellar turbulence \citep{2004RvMP...76..125M,2007prpl.conf...63B} both in the diffuse interstellar medium (ISM), e.g. neutral hydrogen (\ion{H}{1}),  and in the molecular gas  \citep{1994ApJ...427..987B,2007AAS...211.9207H,2001ApJ...546..980O} which is distributed over an extensive range of density regimes in galaxies. While most of the sightlines towards Giant Molecular Clouds (GMCs) have low column density ($\leq10^{22}\,cm^{-3}$), it is the properties of the dense gas components, such as magnetic field strength, turbulent energy, and angular momentum,  that regulate the region where new stars are being born \citep{2010ApJ...709..191M,2008AJ....136.2782L,2015AAS...22511002C,2005AAS...20716302L,2017A&A...603A..64S,2017A&A...607A...2S}. Thus, an exploration of their contribution on the dynamical evolution of molecular clouds is crucial to fully understand the process of star formation \citep{2012MNRAS.421.3522H,2004ApJ...606..271G,1988ApJ...334..613S}. 

Many methods of tracing the magnetic fields have been proposed, although each method has limitations and biases. For instance, measuring polarization from dust grain alignment \citep{2015ARA&A..53..501A,2007JQSRT.106..225L,2015A&A...576A.106P} requires high sensitivity either large scale far infrared or sub-mm polarization maps, which are extremely time-consuming from ground-based telescopes and often have low dust grain alignment efficiency \citep{2007JQSRT.106..225L}. Surveys of magnetic fields from polarization due to the selective extinction of starlight passing through dust clouds are not highly effective for clouds with large dust columns. The synchrotron polarization method \citep{2011piim.book.....D,2002ARA&A..40..319C,2012ApJ...757...14J} is mostly used to trace magnetic field in warm and hot phases of ISM, while though Faraday rotation one can only measure the magnetic field along the line of sight towards regions where the ionization is significant \citep{2013ApJ...777...55H}.

Both numerical simulations and observations have shown that the Velocity Gradient Technique (VGT) is a promising method in studying magnetic fields \citep{2017ApJ...835...41G,2017ApJ...837L..24Y,2018ApJ...853...96L, 2018ApJ...865...46L,2019ApJ...873...16H,survey, PCA,HRO,2019MNRAS.486.4813Z}. The first suggestion to use velocity gradients to trace magnetic fields was made in \cite{2017ApJ...835...41G} (henceforth GL17) where velocity centroid gradients (VCGs) were used as proxies of velocity gradients. This technique was applied to a wide range of column densities from diffuse neutral hydrogen (H I) gas in \cite{2017ApJ...837L..24Y} (henceforth YL17) and \citet{2018MNRAS.480.1333H}, and later \cite{2018ApJ...853...96L} proposed to use velocity channel gradients (VChGs) to trace magnetic fields. The application of the techniques to the diffuse \ion{H}{1} data has proven the promise of this radically new way of magnetic field study \citep{2018ApJ...853...96L}.  A recent exploration of the gradients technique in self-absorbing molecular gas data and self-gravitating media \citep{2017arXiv170303026Y,2017arXiv170303035G,2019ApJ...873...16H,survey,PCA,HRO} have also demonstrated that VGT can be used as tracers of the magnetic field in regions with different physical conditions using $^{13}$CO as molecular tracer.

\begin{table*}
\centering
\begin{tabular}{| c | c | c | c | c | c | c | c |}
\hline
\label{tab:tracer} 
 Molecular Line & Line of Transitions & Frequency [GHz] & FWHM [arcsec] & $\xi [kms^{-1}]$ & $\delta v[kms^{-1}]$ & Tracing Density & $\sigma_{T_R}$[K]\\ \hline \hline
 $^{12}$CO & J = 1 - 0 & 115.27 & 27.12 & 0.18 & 3.48 & $\sim 10^{2}cm^{-3}$ & 0.113\\
 $^{13}$CO & J = 1 - 0 & 110.20 & 28.37 & 0.18 & 4.75 & $\sim 10^{3}cm^{-3}$ &  0.053 \\ 
C$^{18}$O & J = 1 - 0 & 109.78 & 36.01 & 0.18 & 8.32 &$\sim 10^{4}cm^{-3}$ & 0.053 \\\hline
CS & J = 1 - 0 & 48.99 & 63.82 & 0.21 & 9.61 &$\sim 10^{4}cm^{-3}$ & 0.095 \\ 
HNC & J = 1 - 0 & 90.66 & 34.48 & 0.22 & 4.97 & $\sim  10^{5}-10^{6}cm^{-3}$ &  0.039 \\
HCO$^{+}$ & J = 1 - 0 & 89.19 & 35.05 & 0.23 & 4.79 & $\sim  10^{5}-10^{6}cm^{-3}$ & 0.018 \\
HCN & J = 1 - 0 & 88.63 & 35.28 & 0.23 & 6.41 & $\sim  10^{5}-10^{6}cm^{-3}$ & 0.019\\ \hline
\end{tabular}
\caption{\label{tab:tracer1} Observation parameters for seven chemical tracers used in our analysis, including the type of molecules, their respective transition levels,  observing frequency in GHz, original telescope beam full width half maximum (FWHM) without any additional smoothing in arcsec, the smallest thickness of the velocity channel slice, and the approximate molecular hydrogen density (N$_{H_2}$) we expect each molecule to trace. Note that for all of molecules we have observed the ground state (J\,=\,1\,--\,0) transitions. $\xi$ is the velocity channel width of each molecular line cube.  Note that the cubes are Nyquist sampled so the true velocity resolution is 2$\xi$. $\delta v$ is the velocity dispersion and $\sigma_{T_R}$ is the per channel noise level of $T_R$ for each data cube. \cite{2018arXiv180408979F} provides more details about the data reduction.}
\end{table*}
In this paper, we present an observational example for a low galactic latitude Giant Molecular Cloud  Vela C using the recent advancement of the VGT made by \cite{2019ApJ...873...16H} in numerical simulations, which demonstrated that VGT can also be applied to CO isotopologues including $^{12}$CO and C$^{18}$O. We use 7 molecular line maps for the analysis in Vela C, including the 3 isotopes of CO, CS, HNC, HCO$^+$, and HCN. The latter four molecular tracers are high density tracers (number density of H$_2$ around $10^{4}$\,cm$^{-3} - 10^{6}$\,cm$^{-3}$) compared to the CO tracers $^{12}$CO, $^{13}$CO, and C$^{18}$O, which typically trace number densities of H$_2$ between $10^{2}$\,cm$^{-3} - 10^{4}$\,cm$^{-3}$ \citep{2015PASP..127..299S,2018arXiv180408979F}.  To evaluate the success of the VGT we compare our inferred magnetic field orientation from the VGT to the magnetic field orientation inferred from a large scale 500\,$\mu$m BLASTPol polarization map first presented in \cite{2016ApJ...824..134F}.

In what follows, in Section \ref{sec:theory} and Section \ref{sec.method}, we describe the theoretical foundation and the Velocity Gradients Technique used in our work. In Section \ref{sec:co}, we discuss the ability of VGT to trace magnetic fields over a large range of densities in GMCs and the contribution from the foreground and background. In Section \ref{sec:surrounding and collapsing}, we show how to trace the magnetic field through the combination of different molecular tracers and estimate the fraction of the collapsing regions in Vela C. In Section \ref{sec:discussion} we discuss the possible application of gradient technique to other molecular clouds. In Section \ref{sec:conclusion}, we give our conclusions.

\section{Theoretical Perspective on Self-Absorbing Gradient Technique}
\label{sec:theory}
\subsection{Basic MHD Turbulence Theory}

The MHD turbulence has been explored both theoretically and numerically in decades \citep{1983JPlPh..29..525S,1984ApJ...285..109H,1995ApJ...447..706M}. In \cite{1995ApJ...438..763G} (henceforth GS95) they formulated the theory of incompressible MHD turbulence and also predicted predicted the turbulent anisotropy which later becomes the foundation of several magnetic fields tracing techniques. GS95 described that the scaling of turbulent eddies is approximate $v_{l}\sim l^{\frac{1}{3}}$, where $v_{l}$ is the turbulence velocity at scale $l$ and $l$ is the size of eddies perpendicular to the magnetic field. However, GS95 is done with respect to the mean magnetic field where the anisotropic relation is not expected to be observed. 

In \citet{1999ApJ...517..700L} they illustrate that the motion of turbulent eddies enables the magnetic fields to mix with minimal resistance from magnetic tension for eddies at all scales and thus a much faster rate of magnetic reconnection is allowed compared to the traditional Sweet-Parker model. As a result of the mixing motions, the Alfv\'{e}nic turbulence tends to move along the magnetic field directions. To find how turbulent eddies evolve in the direction parallel to the magnetic field, it is also necessary to consider the mixing motions associated with magnetic eddies and Alfv\'{e}n waves with the period equal to the period of an eddy:
\begin{equation}
\frac{l}{v_{l}} \sim \frac{l_{||}}{v_{A}}   
\end{equation}
where $l_{||}$ is the parallel scale of the eddy and $v_{A}$ is Alfv\'{e}n velocity. The correlation between the parallel and perpendicular scales of sub-Alfv\'{e}nic turbulence, i.e. $v_{l}\le v_{A}$,  eddies can be obtained as:
\begin{equation}
l_{||}\sim l^{\frac{2}{3}}
\end{equation}
This correlation shows that turbulent eddies are elongated along the direction of the magnetic field and it holds for the eddies that are aligned with the local direction of the magnetic field that surrounds them. Incidentally, the concept of local magnetic field frame explains why, unlike the original GS95 treatment, the anisotropy of turbulence actually reflects the direction of magnetic fields that percolates turbulent eddies. This result was confirmed by numerical simulations \citep{2000ApJ...539..273C,Cho2003CompressibleImplications,2001ApJ...554.1175M}. Due to this particular property of MHD turbulence, the velocities associated with a turbulent eddy are anisotropic so that the largest change of the velocity is in the direction perpendicular to the local direction of the magnetic field. Thus, it is essential that \textbf{VGT is tracing the local magnetic field around eddies rather than the mean magnetic field.}

It is worth to mentioning that several approaches have been proposed to trace the magnetic field based on the MHD anisotropy relation Eq.~(2). The correlation function analysis (CFA) of the velocities and the Principal Component Analysis for Anisotropy (PCAA) were first proposed to study magnetic field morphology \citep{2008ApJ...680..420H}, to estimate magnetization \citep{2015ApJ...814...77E,2011ApJ...740..117E}, and to determine the contribution of the fast, slow and Alfven modes in observed turbulence \citep{2016MNRAS.461.1227K,2017MNRAS.464.3617K,2017MNRAS.470.3103K} using the theoretical understanding of MHD turbulence discussed in Section 2.1. However, numerical study in \citet{2018ApJ...865...54Y} showed that comparing with VGT, CFA and PCAA face several issues in particular, the anisotropy may be distorted, multi-centered eddies or the contours are not closed. These significantly degrade the determination of the direction of anisotropy, thus the inferred magnetic field orientation through CFA and PCAA. Later \citet{2014ApJ...789...82C} proposed the Rolling Hough Transform (RHT) to study the magnetic field in the diffuse region based on the fact that so-called narrow "\ion{H}{1} fibers" are align parallel to the magnetic field orientation. However, the RHT requires linear structures in the ISM \citep{2014ApJ...789...82C}.

Moreover, VGT as a superior technique has been successfully tested to trace the local magnetic field from both diffuse region and absorbing media for the case of $^{13}$CO emission with different abundances and densities \citep{2017arXiv170303035G}, while \citet{2019ApJ...873...16H} showed in numerical simulations that the VGT can map out the magnetic field structures of different physical depths in a molecular cloud using CO isotopologues with different optical depths and in the presence of weak self-gravity. 

\subsection{Molecules as Probes of Gas in Different Density Regimes}

Giant Molecular Clouds (GMCs) are the sites for most of the star formation to take place across the Milky Way and also in other galaxies. The star formation regions are often observationally unresolved in GMCs. The gas in GMCs have an average number density of the order of $10^{2}cm^{-3}$ and temperature of 10 - 30K \citep{2009ApJ...699.1092H}, while the star formation clumps may have number densities around to $10^{7}cm^{-3}$ and temperature as low as 10K or even lower \citep{1994ApJ...428..693W}. Molecular tracers like $^{12}$CO typically become optically thick, and therefore mostly trace the outer (low density) regions of molecular clouds.
$^{13}$CO, which is typically a factor of $\sim$100 less abundant than  $^{12}$CO and therefore usually optically thin, can be used to trace intermediate density regions, as can the even less abundant isotopologue C$^{18}$O. Transitions of the molecules CS, HNC, HCO$^{+}$  and HCN have higher critical densities and are therefore used to trace intermediate or high-density molecular gas. These molecules tend to either trace regions where the effect of self-gravitation is very strong or gas that are just outside the self-gravitating regions. However, regions with high-density gas do not necessarily host star formation, that is, high density does not necessarily indicate that the self-gravitation is strong enough to trigger the gravitational collapse. 

In summary, we expect that each molecular line will trace a different set of temperature, density, and excitation conditions. Therefore, by using multiple molecular lines it is possible to study the cloud velocity and density structure in low, intermediate and high-density molecular gas. Table \ref{tab:tracer1} summarizes the properties of the molecular lines used in this study. Note that we only use the ground state ($J\,=\,$1\,--\,0) transition of each molecular line in our analysis.
\section{Method}
\label{sec.method}
\subsection{Velocity Gradients Technique}
We use molecular line data from a survey with the 22-meter Mopra telescope covering the Vela\,C Giant Molecular Cloud, covering 264.62$^\circ$ - 266.56$^\circ$ in galactic longitude, and 0.54$^\circ$ - 1.92$^\circ$ in galactic latitude. The observations are is described in detail in \cite{2018arXiv180408979F}. The beam full width half maximum (FWHM) in arcmin and the thickness of velocity slice in $kms^{-1}$ of different tracers are shown also in Table \ref{tab:tracer}.

Vela\,C is a massive ($M\,\sim\,10^5\,M_{\sun}$),  relatively nearby GMC (GAIA-DR2 distance $\sim$\,900\,pc), that appears to be relatively young and unevolved \citep{1999PASJ...51..775Y,2009ApJ...707.1824N,2011A&A...533A..94H}.  This molecular cloud is mostly cold ($T_{\mathrm{dust}}\,<\,$15\,K) and in early stages of star formation, though it has formed a 1\,Myr cluster of stars which powers a compact \ion{H}{2} region RCW 36 \citep{2013A&A...558A.102E}. Since Vela\,C is dominated by a single velocity component in the range 0 - 12 km/s \citep{2018arXiv180408979F}, it therefore provides a good opportunity for testing the Velocity Gradient Technique through the study of alignment between VCGs/ VChGs and the magnetic field orientation inferred from polarization data. We infer the orientation of the projected magnetic field by rotating measurements of polarization angle made with the Balloon-borne Large-Aperture Sub-millimeter Telescope for Polarimetry (BLASTPol) at 500\,$\mu$m by 90$^{\circ}$ \citep{2016ApJ...824..134F}.\footnote{ The data reduction and removal of the contribution to the polarized emission from dust in the diffuse ISM surrounding Vela\,C is described in \cite{2016ApJ...824..134F}. In our analysis, we use only data using the "Intermediate" diffuse emission subtraction method from \cite{2016ApJ...824..134F}. The beam FWHM of the BLASTPol 500$\mu$m polarization data is 2.5\arcmin, which corresponds to $\sim$0.6\,pc at the distance to Vela\,C.}

Velocity centroid maps \textbf{C(x,y)} for all tracers are produced by integrating along the velocity axis of the PPV (Position-Position-Velocity) cube:
\begin{equation}
C(x,y)=\frac{\int dv T_{R}(x,y,v)\cdot v }{\int dv T_{R}(x,y,v)}
\end{equation}
where $T_{R}$ is the the radiation temperature (in units of Kelvin), and v is the line-of-sight velocity.
\begin{figure*}
\centering
\includegraphics[width=0.98\linewidth,height=0.5\linewidth]{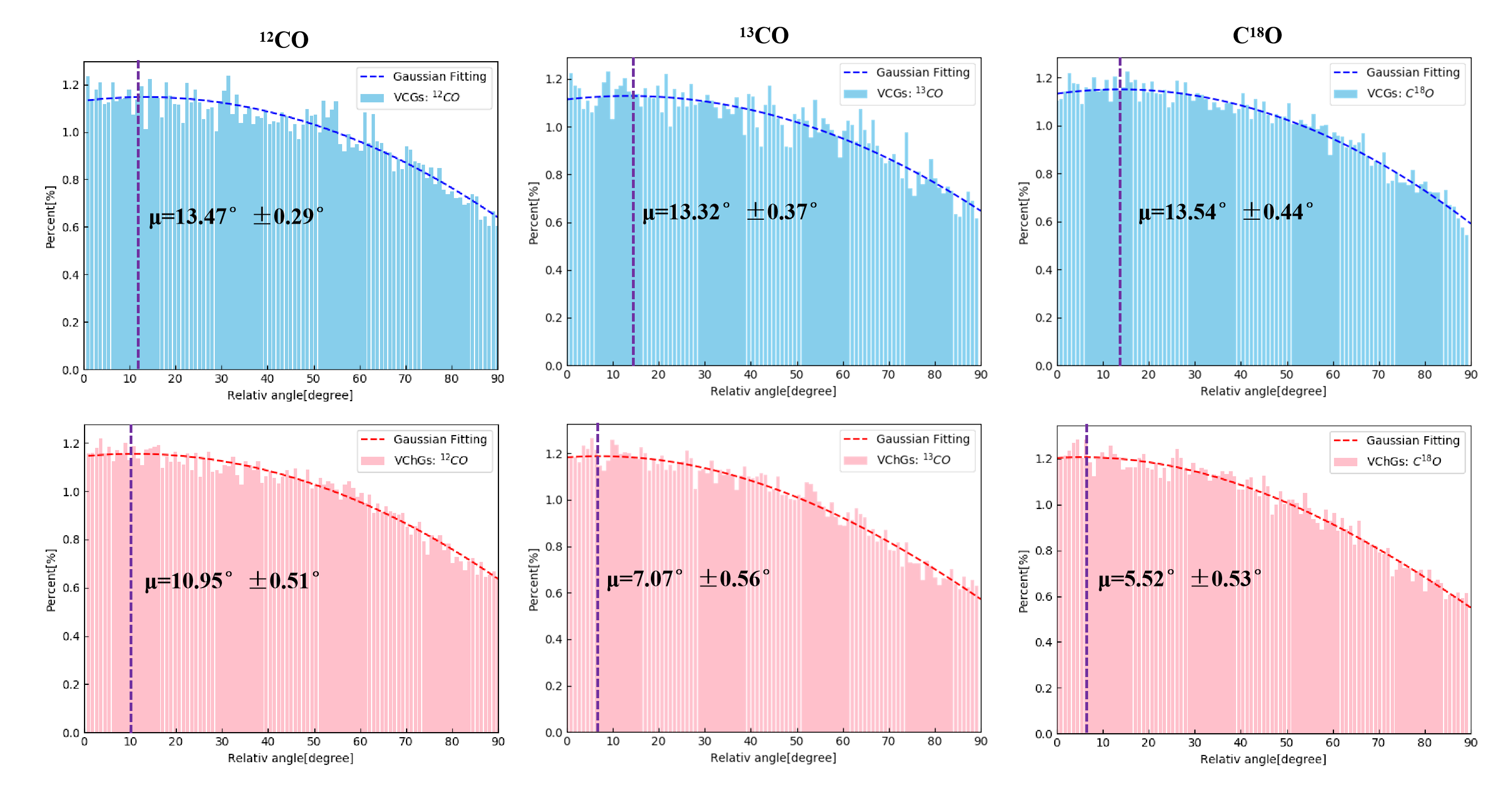}
\caption{\label{fig:ad_co}Normalized distributions of alignments between rotated Velocity Centroid Gradients (VCGs, top row) / Velocity Channel Gradients (VChGs, bottom row) and magnetic field inferred from polarization. The distribution is drawn by using raw gradients of each pixel without sub-block averaging. The dashed line is a Gaussian fit to the distribution, where
$\mu$ is the expectation of the distribution. The vertical axis is the relative probability of the relative angle. The uncertainty is given by the standard error of the mean, i.e. the standard deviation divided by the square root of the sample size, while the systematic uncertainty in polarization data is given in Appendix ~\ref{appendx:A}.}
\end{figure*}
\begin{figure*}
\centering
\includegraphics[width=0.99\linewidth,height=0.50\linewidth]{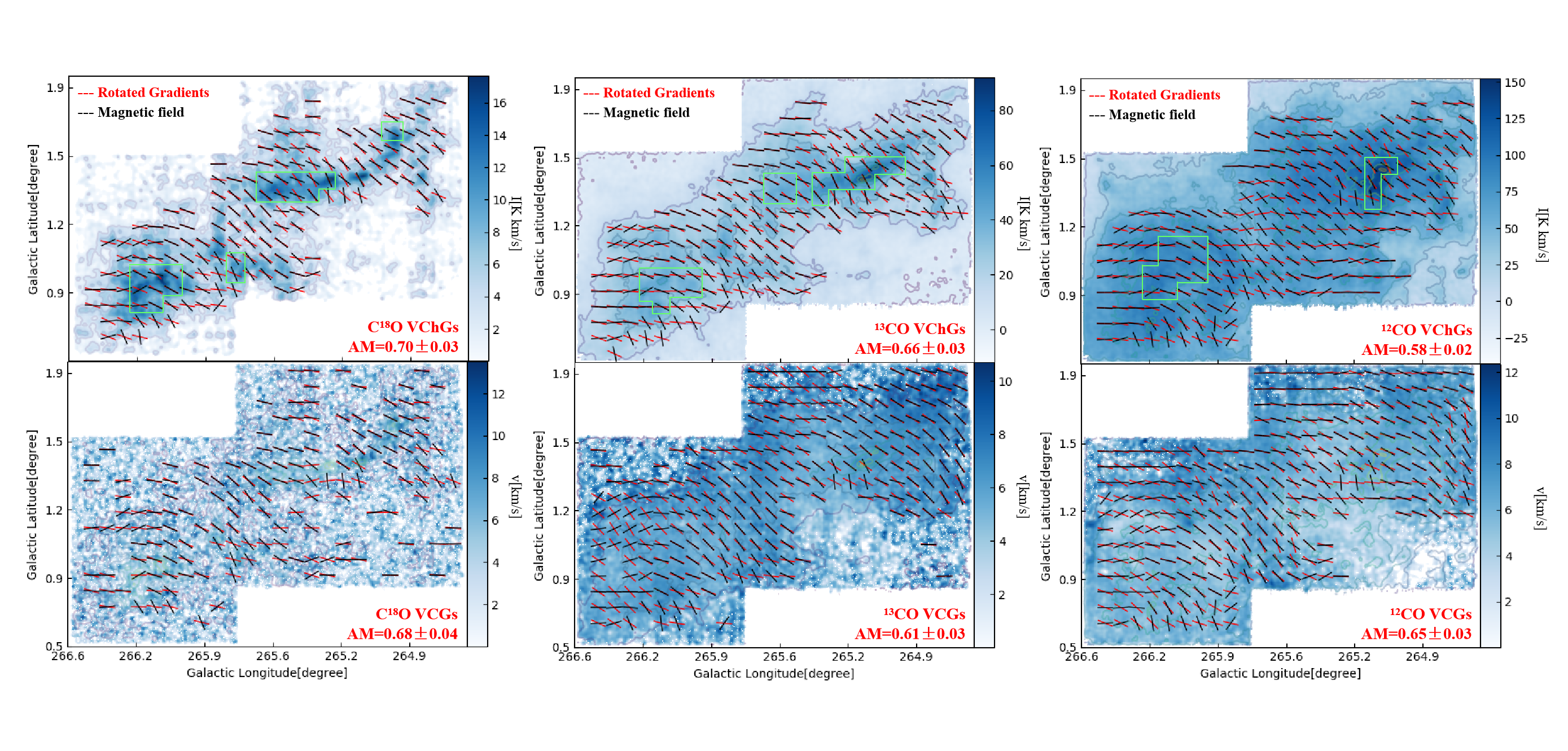}
\caption{\label{fig:CO}The orientation of VChGs (top row), VCGs (bottom row), and the magnetic field obtained from polarization (black line segment). The VCGs and VChGs are rotated by 90$^\circ$. The left panels use C$^{18}$O as a tracer, the middle panels show $^{13}$CO, the right panels show $^{12}$CO. The background images are moment zero maps (integrated line intensities) for VChGs, and the first moment maps for the VCGs. Green contours in the top panels indicate the high-intensity regions in which all pixels are above the 95$^{th}$ percentile in integrated line intensity. (See \citet{2018arXiv180408979F} for details of each data set.)}
\end{figure*}
For the analysis of observational data, the theory describing the statistics of fluctuations in the Position-Position-Velocity (PPV) cube is crucial \citep{2000ApJ...537..720L}. It predicts the anisotropy of the velocity channel maps \citep{2002ASPC..276..182L}. 

Even when the velocity slice is thin, the channels can record more contribution from turbulent velocities \citep{2000ApJ...537..720L,comment}. Therefore, the gradients of thin velocity channel maps (VChGs) are expected to trace the magnetic field orientation with high accuracy \citep{2018ApJ...853...96L}.  Thus, we also investigate the correspondence between VChGs and the magnetic field. To construct the velocity channel maps, we create integrated maps over a narrow velocity range $\Delta v$ satisfying:
\begin{equation}
\Delta v \leq \sqrt{\langle\delta v^{2}\rangle}
\end{equation}
where $\delta v^{2}$ is the line of sight velocity dispersion. This is the criterion proposed in \citet{2000ApJ...537..720L} so that the velocity contribution in the velocity channel map dominates over the density contribution (See Tab~\ref{tab:tracer}). Then the velocity channel map can be calculated also by integrating along the velocity axis:
\begin{equation}
Ch(x,y)=\int_{v_0-\Delta v/2}^{v_0+\Delta v/2} dv\ T_R(x,y,v)
\end{equation}

where $v_0$ is the velocity corresponding to the central peak of the velocity profile along the line-of-sight. We follow the gradient calculation algorithm of velocity centroid maps from \cite{2017ApJ...837L..24Y}, and velocity channel maps from \cite{2018ApJ...853...96L}, with the sub-block averaging method \citep{2017ApJ...837L..24Y} applied (Note that the sub-block averaging method is not just a smoothing method for suppressing noise in a region, but used to increase the reliability of the important statistical measure.). In this work, the sub-block size is selected as 24 pixels ($\sim 1pc$, 12 arcsec per pixel for CO isotopologues, HCN, HNC, HCO$^{+}$, while 24 arcsec per pixel for CS) which is enough for statistical accuracy. The polarization vector is averaged over the same sub-block size. In addition, \cite{2018ApJ...853...96L} suggested a refinement of the velocity gradient technique by using the Moving Window approach which is using a continuous sub-block averaging, rather than dividing up the map into discrete sub-blocks. We apply the Moving Window to all our gradient maps unless specifically mentioned.

In practical observation, the noise is present in spectroscopic data. \cite{2018ApJ...853...96L} find that the white noise would alter the alignment of gradients to the projected magnetic field. However, the alignment remains fairly good for data with intensity signal to noise of 3 or greater, and only significantly diverges from the true gradient alignment angle when the signal to noise of the data approaches 1. Therefore, in the case of high noise situation, a Gaussian filter is proposed in \cite{2017ApJ...842...30L} to reduce the effect of noise in velocity centroid maps and velocity channel maps.  In addition, in order to match the gradient after the sub-block averaging, the polarization data is firstly smoothed to the same angular resolution as molecular data We average the polarization data over the area as the sub-block size in order to match the resolution of the gradient maps. With a sub-block size 24 pixels, the effective resolution of both gradients and polarization is 4.8' (9.6' for CS)  which is larger than the minimum resolution of polarization 2.5'. Note that neither the polarization nor the spectroscopic data are filtering out scales smaller than  4.8' (9.6' for CS), and there could be small-scale magnetic field structures that cannot be resolved with our sub-block averaging method.

\subsection{Alignment Measure}
The orientation of gradients from VGT is compared with the magnetic filed inferred from BLASTPol polarization. The relative orientations between the 90$^\circ$ rotated gradients and project magnetic field directions from polarization angles are measured by the \textbf{Alignment Measure (AM)} used: 
\begin{align}
AM=2(\langle cos^{2} \theta_{r}\rangle-\frac{1}{2})
\end{align}
where $\theta_r$ is the angular difference between the gradient vector rotated by 90$^\circ$ and the magnetic field vector derived from polarization in a single sub-block. $\langle...\rangle$ indicates the average over all sub-blocks.

The range of AM is [-1,1]. There are three important cases for the value of AM:
\begin{enumerate}
\item AM = -1: in this case the gradients are perpendicular to the projected magnetic field; 
\item AM = 0:  the gradients are neither perpendicular nor parallel to the magnetic field;
\item AM = 1: the gradients are parallel to the projected magnetic field, which implies a perfect alignment. 
\end{enumerate}
The AM is a parameter used to quantify the relative orientation between two vectors. Note that since AM is insensitive to the sign of the velocity gradient, e.g.~AM($\theta$)\,=\,AM($-\theta$) , it is advantageous to test the performance of the VGT  in terms of the orientation of the plane-of-the-sky magnetic fields, i.e. the larger AM, the better alignment between the measurement from VGT and polarization.

\begin{figure*}
\centering
\includegraphics[width=0.98\linewidth,height=0.45\linewidth]{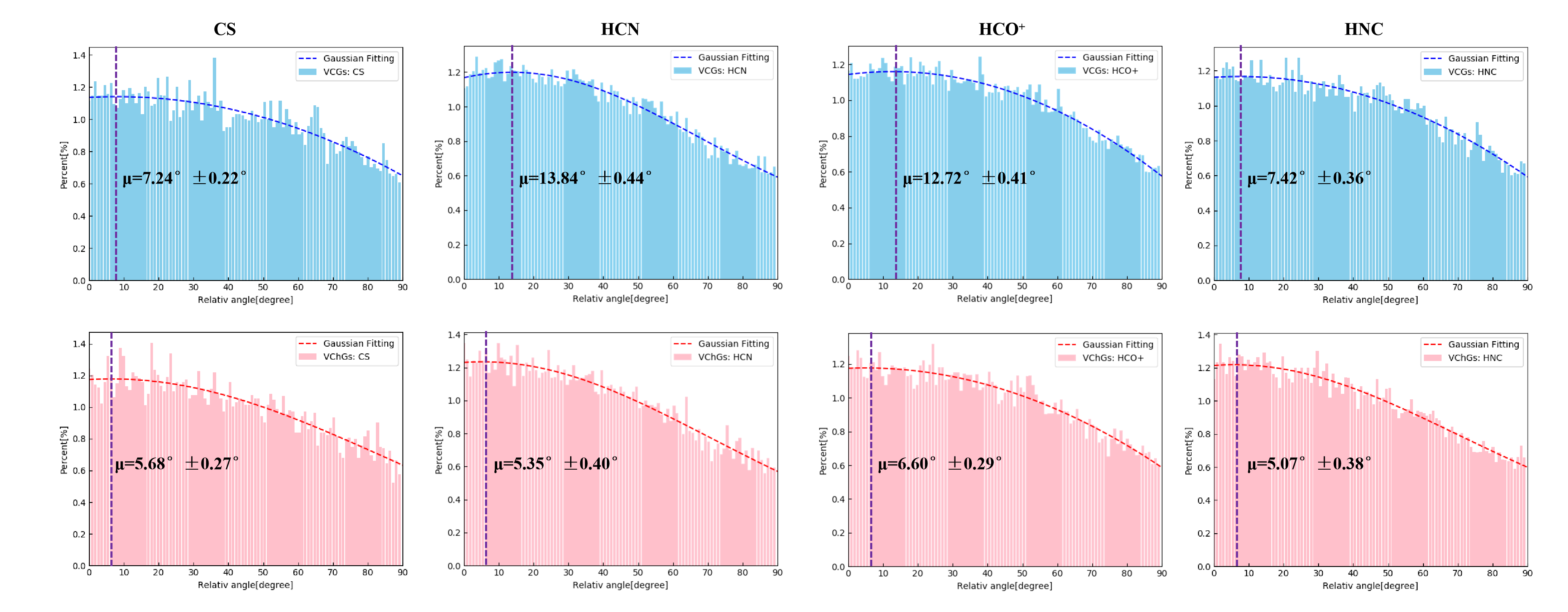}
\caption{\label{fig:ad_cs}Normalized distribution of alignment between rotated VCGs (top row) / VChGs (bottom row) and the magnetic field inferred from polarization. The distribution is drawn by using raw gradients of each pixel, without sub-block averaging. The dashed line is a Gaussian fit to the distribution, where
$\mu$ is the expectation of the distribution. The vertical axis is the relative probability of the relative angle. The uncertainty is given by the standard error of the mean, i.e. the standard deviation divided by the square root of the sample size, while while the systematic uncertainty in polarization data is given in Appendix ~\ref{appendx:A}.}
\end{figure*}
\begin{figure*}
\centering
\includegraphics[width=0.99\linewidth,height=0.5\linewidth]{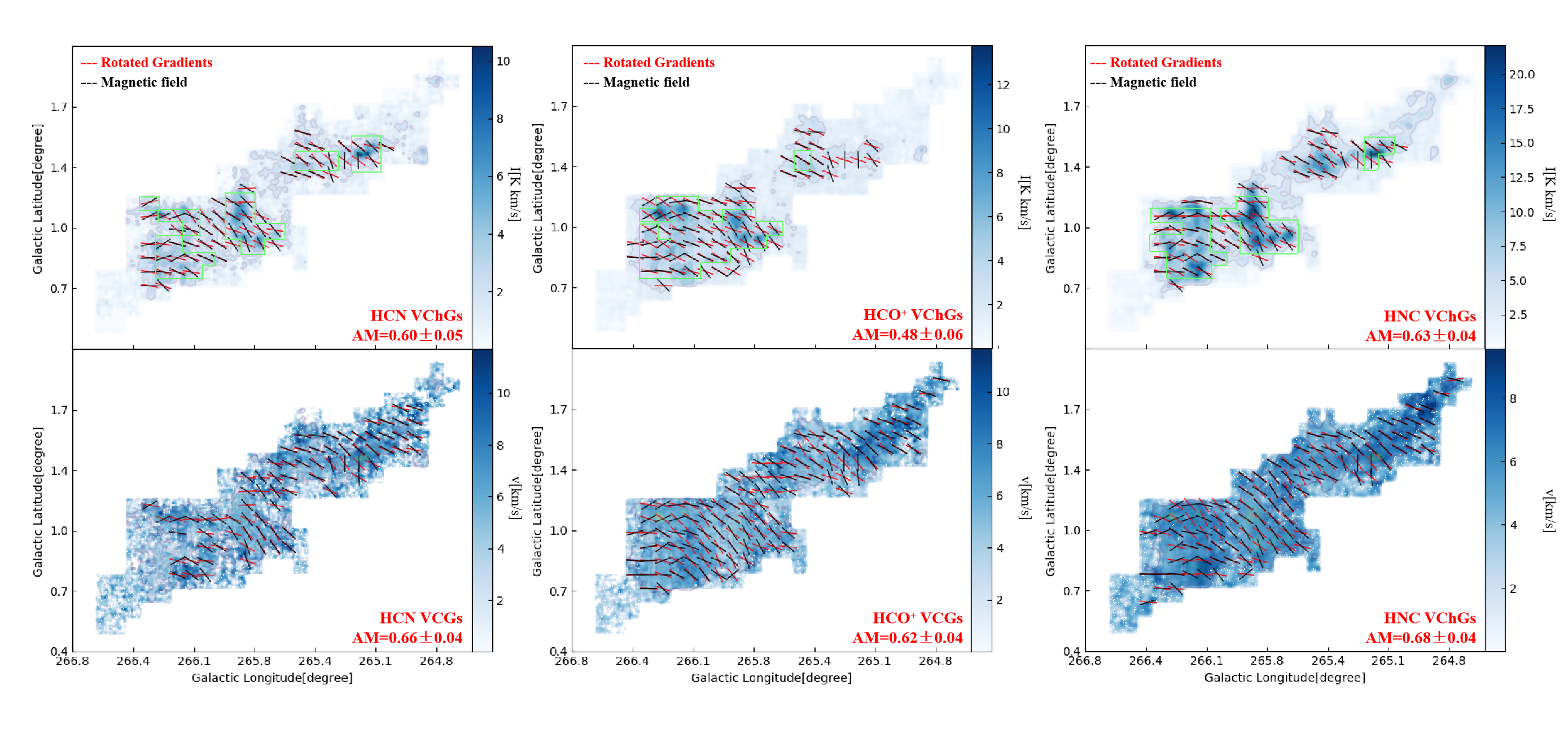}
\caption{\label{fig:HNC}The orientation of VChGs (top row), VCGs (bottom row), and the magnetic field (black line segment) obtained from polarization. The VCGs and VChGs are rotated by 90$^\circ$. The left column represents using HCN as tracer, the second column is HCO$^{+}$, the third column is HNC.  Green contours in the top panels indicate the high-intensity regions in which all pixels are above the 95$^{th}$ percentile in integrated line intensity. (See \citet{2018arXiv180408979F} for details of each data set.)}
\end{figure*}
\section{Tracing the Magnetic Field using VGT over a larger range of densities}
\label{sec:co}
\subsection{Low-density tracers: $^{12}$CO, $^{13}$CO \& C$^{18}$O}
$^{12}$CO, $^{13}$CO, and C$^{18}$O typically trace gas with H$_2$ number density between  $10^{2}cm^{-3}-10^{4}cm^{-3}$, which is the typical density for a young self-gravitating molecular cloud \citep{2012ARAA...50..29}. Because their lines tend to have lower optical depths, $^{13}$CO and C$^{18}$O can trace molecular gas over a larger range of densities, while optically thick $^{12}$CO traces the lower density outer cloud regions. In \cite{2017ApJ...837L..24Y}, we see that by combining the gradient vectors and polarization measurements from the same region, we can determine whether a region is optically thick or thin, and whether the self-gravity is dominant in a particular region or not. 

Fig.~\ref{fig:ad_co} shows the normalized distribution of the relative orientation between rotated VCGs/VChGs and the inferred magnetic field from polarization using $^{12}$CO, $^{13}$CO, and C$^{18}$O as tracers. The distribution is constructed by calculating gradients for each map pixel, without using sub-block averaging. We find that the set of rotated gradient alignment angles is roughly consistent with a Gaussian distribution. The expectation values of VCGs are located near 13.5$^o$, while for VChGs the expectation values are decreasing with the increasing of the critical density of each tracer. Considering the systematic uncertainty in polarization data is less than 10$^\circ$ (see Appendix~\ref{appendx:A}), the results support our theoretical consideration that the velocity gradient rotated by 90$^\circ$ tends to align with the magnetic field.

Fig.~\ref{fig:CO} shows the orientation of VCGs, VChGs, and the magnetic field obtained from polarization with sub-block averaging method applied. Fig.~\ref{fig:CO} also shows that using  C$^{18}$O VChGs reflects the central structure of the cloud, while $^{12}$CO and $^{13}$CO determine the outskirts structure. Considering the systematic uncertainty in polarization data is approximately 2.07$^\circ$ (see Appendix~\ref{appendx:A}), these three tracers shows similar agreements with the magnetic field inferred from polarization data based on the estimation of AM.

\subsection{High-density tracers: HCN, HNC, HCO$^+$ \& CS}
\label{sec:cs}
Molecular CS is well known for tracing dense clumps
in the molecular clouds in the Milky Way. Also, recent
multi-line surveys of nearby spiral galaxies \citep{2009ApJ...707..126B} of this molecule in extragalactic environments have revealed low J (J $\leq$ 4) transitions of CS trace gas densities of the order $10^{5}cm^{-3}$, even the number is close to $10^{6}cm^{-3}$. Estimates of the effective excitation density of CS from \citet{2015PASP..127..299S} show that cold CS J\,=\,1\,--\,0 typically traces densities of $\sim$10$^4$\,cm$^{-3}$.

HCN and HNC are two species that have routinely been used as tracers of star formation regions in molecular clouds \citep{1997ApJ...483..235T}. In fact, HCN is used to trace the same approximate densities as the low-J CS molecule, and hence the gas that it traces is not necessarily in a self-gravitating star formation region. HNC has a similar critical density to that of HCN. Therefore, HNC and HCN should be good tracers for higher density star formation gas which might be self-gravitating. \citet{2017A&A...599A..98P} showed that HNC is a better molecular tracer than HCN when probing low-J lines, based on the analysis of visual extinction and line integrated intensities. We note that the HCN J\,=\,1\,--\,0 has hyperfine structure which complicates the interpretation of the VCGs.  Also the HCN emission has a lower integrated line emission compared to HNC \citep{2018arXiv180408979F}.

Compared to the other three tracers, HCO$^{+}$ is more often to be used as a tracer of ionized gas (e.g., see \citealt{2015ApJ...799..204C}). However, 80\% of the HCO$^{+}$ and HCN emission originates in non-self-gravitating regions of molecular clouds \citep{2017A&A...602A..51V}. HCN and the HCO$^{+}$ emission lines can therefore both be used to trace the dense gas around the self-gravitating regions.
\begin{figure}
\centering
\includegraphics[width=0.95\linewidth,height=1.25\linewidth]{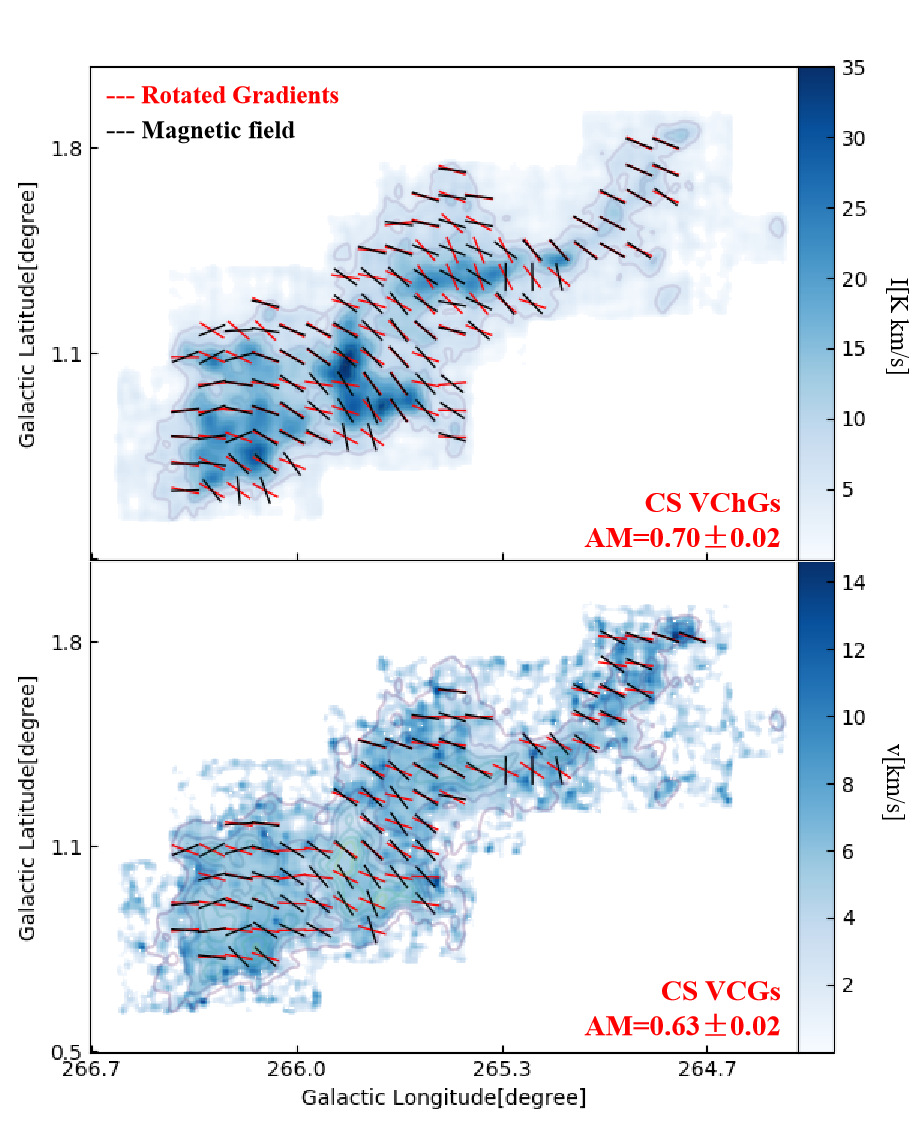}
\caption{\label{fig:CS}The orientation of VChGs (top row), VCGs (bottom row), and the magnetic field (black line segment) obtained from polarization. The VCGs and VChGs are rotated by 90$^\circ$.}
\end{figure}

Fig.~\ref{fig:ad_cs} shows the normalized distribution of the relative orientation between rotated VCGs/VChGs and the inferred magnetic field from polarization, using CS, HCN, HNC, and HCO$^{+}$. We do not apply sub-block averaging when constructing the distributions, but use the raw gradients for each pixel. We find that the distribution is consistent with a Gaussian distribution. However, the expectation values of VChGs are $\sim$5.5$^{\circ}$, lower than the expectation values for the $^{12}$CO, $^{13}$CO, and C$^{18}$O lines. CS and HNC show smaller expectation values of VCGs compared to HCN and HCO$^+$.

Fig.~\ref{fig:HNC} and Fig.~\ref{fig:CS} show the orientation of VCGs, VChGs using HCN, HNC, HCO$^{+}$, and CS, and the magnetic field obtained from polarization  with sub-block averaging method applied. Fig.~\ref{fig:CS} shows that CS is able to detect a clear structure of Vela C by using VChGs, and shows perhaps the best agreement with the polarization inferred magnetic field orientation, with an AM=0.70. Among four tracers, HNC and CS appear to best trace the magnetic field.

\subsection{Contribution from the Foreground and Background}
In a low galactic latitude cloud, such as Vela\,C, foreground and background material can change the observed cloud magnetic field orientation. In contrast, the VGT method for tracing magnetic fields is only sensitive to the molecule probed, and molecular line cubes can be studied solely over the velocity range associated with the molecular cloud of interest. Using the VGT it should therefore also be possible to trace the magnetic field in different layers of the cloud, by targeting lines that trace different ranges of densities. Thus, VGT using molecular tracers provides the information of the localized magnetic field corresponding to species in the cloud, while polarization accumulates the information along the line of sight, i.e. VGT and polarization are tracing different components of the magnetic field. If we want to fairly compare VGT and polarization, we should consider the contribution from the foreground and background\footnote{ Note that \citet{2018arXiv180408979F} shows that Vela C is dominated by a single velocity component in the range 0 - 12 km/s, while there might be multiple velocity components in a wider velocity range. For a fair comparison with polarization, we did not constraint the PPV cube to a particular velocity range, but keep its original velocity range -20 - 30 km/s for the presented results. We repeat our VCGs and VChGs analyses but limit the velocity range as 0 - 12km/s. The AM (with error bar $\sim \pm0.03$) is generally 0.05 lower than the presented results. AM values for single/multiple component analyses still give similar values. We, therefore, expect the bulk of the signal in both data can be reasonably well compared and is dominated by signals from Vela C, although both polarization and spectroscopic data are not entirely constrained to Vela C.}

In \citet{2016ApJ...824..134F} the authors attempted to remove the contribution of the foreground and background dust to the BLASTPol 500$\,\mu$m polarization data by using two different methods to model the diffuse polarized dust emission, and subtracted their model from the data.  However the systematic uncertainty associated with removing the diffuse polarized emission is expected to most affect polarization angles towards regions of low column density within Vela C.  We note that several locations where the BLASTPol inferred magnetic field direction differs from the velocity gradient inferred field direction are towards the edge of the map in low column density regions of Vela C.
 
Tab.\ref{tab:angle} shows the expectation value $\mu$ of the relative angle between the rotated gradients of each molecular tracer and the magnetic field inferred from dust polarization.  From Table.\ref{tab:angle}, we find that low-density tracers, i.e. $^{12}$CO and $^{13}$CO, indeed show larger deviation between the rotated VChGs and magnetic fields inferred from the polarization than high-density tracers. As mentioned above, for low-column density regions we expect a larger systematic uncertainty in the polarization angle  associated with background/foreground polarized emission subtraction, so it is not surprising that $^{12}$CO and $^{13}$CO, which show emission across the entire mapped area, show a larger deviation from the VChG and VCG inferred magnetic field orientation. 

In addition, we find that VChGs show approximately $\mu\sim6^\circ$ offsets  for dense tracers, which are smaller than the offsets of VCGs. Noise is one possible factor contributing to the larger deviations of the VCGs compared to VChGs, for example, the map of VCGs is integrating over the whole velocity range of the PPV cube, while for VChGs, the map is integrating over a narrow velocity range $\Delta v$. There is also the second possibility for the worse alignment measures for the VCG maps. Although Vela C is dominated by a single velocity component, there are multiple velocity components in some parts of Vela C (see Fig.~1 in \citet{2018arXiv180408979F}). The VCG analysis implicitly assumes only one velocity component. We note also that the HCN J\,=\,1\,--\,0 line has hyperfine structure, which makes calculating line velocity centroids more difficult. In addition, RCW 36 is a compact \ion{H}{2} region on Vela C. However, the \ion{H}{2} region is compact ($\sim$1pc in size) and as such only compromises a small portion of the map \cite{2016ApJ...824..134F}. The contribution from RCW 36 is therefore slim.

\begin{table*}
\centering
\begin{tabular}{| c |c | c | c | c |}
\hline
Molecule Line & $\mu$ (VCGs) & $\mu$ (VChGs) & AM (VCGs) & AM (VChGs)\\\hline\hline
$^{12}$CO & 13.47$^\circ \pm 0.29^\circ$ & 10.95$^\circ \pm 0.51^\circ$ & 0.65$\pm 0.03$ & 0.58$\pm 0.02$\\\hline
$^{13}$CO & 13.32$^\circ \pm 0.37^\circ$ & 7.07$^\circ \pm 0.56^\circ$ & 0.61$\pm 0.03$ & 0.66$\pm 0.03$\\\hline
C$^{18}$O & 13.64$^\circ \pm 0.37^\circ$ & 5.52$^\circ \pm 0.53^\circ$ & 0.68$\pm 0.04$ & 0.70$\pm 0.03$\\\hline
CS & 7.24$^\circ \pm 0.22^\circ$ & 5.68$^\circ \pm 0.27^\circ$ & 0.63$\pm 0.02$ & 0.70$\pm 0.02$\\\hline
HCN & 13.84$^\circ \pm 0.44^\circ$ & 5.35$^\circ \pm 0.40^\circ$ & 0.66$\pm 0.04$ & 0.60$\pm 0.05$\\\hline
HNC & 7.42$^\circ \pm 0.36^\circ$ & 5.07$^\circ \pm 0.38^\circ$ & 0.63$\pm 0.04$ & 0.68$\pm 0.04$\\\hline
HCO$^{+}$ & 12.72$^\circ \pm 0.41^\circ$ & 6.60$^\circ \pm 0.29^\circ$ & 0.48$\pm 0.06$ & 0.62$\pm 0.04$\\\hline
\end{tabular}
\caption{\label{tab:angle}The expectation value $\mu$ of the relative angle between rotated VCGs/VChGs and the magnetic field inferred from polarization, without sub-block averaging method. The uncertainty is given by the standard error of the mean, i.e. the standard deviation divided by the square root of the sample size, while the systematic uncertainty in polarization data is given in Appendix \ref{appendx:A}.}
\end{table*}

\begin{figure}
\centering
\includegraphics[width=0.99\linewidth]{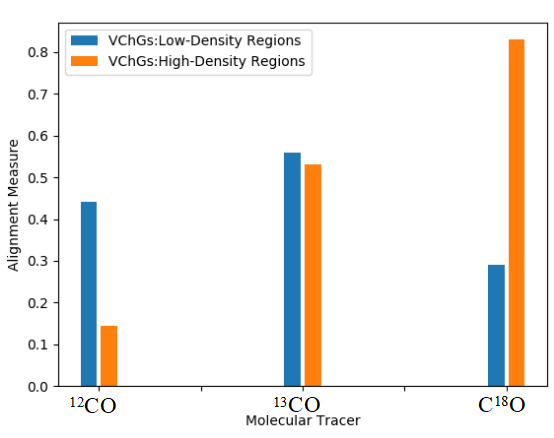}
\caption{\label{fig:density} The alignment measure of low-intensity regions and high-intensity regions as traced by moment = 0 maps (shown in Fig~.\ref{fig:CO}), using VChGs for  molecular tracers $^{12}$CO, $^{13}$CO, and C$^{18}$O.}
\end{figure}
\section{Tracing the Magnetic Field from low-density gas Region to High-density gas Region}
\label{sec:surrounding and collapsing}

\subsection{Low-density Gas Region}
 Fig.~\ref{fig:CO} shows that the integrated velocity channel maps, i.e.~the moment = 0 maps, are able to trace the cloud's column density structure. VChGs of $^{12}$CO and $^{13}$CO appear to mostly trace the low to intermediate density material of the cloud, as they become optically thick towards high column density cloud sightlines.

 Due to the lower optical depth of C$^{18}$O,  its emission lines traces deeper into the cloud, and therefore includes cloud structures over a larger range of densities.  In this case, we can use $^{12}$CO, $^{13}$CO to get a fairly good idea of the magnetic field structure in the outer layers of the cloud and use C$^{18}$O to better trace higher-density structure.

We therefore propose a method that uses multiple molecular lines as a combined tracer for magnetic field in GMCs, even if part of the cloud are strongly self-gravitating. We separate  each moment = 0 map of the Vela C Cloud into the low-intensity regions and the high-intensity regions. The high-intensity region is defined by the percentile of the integrated line intensity: we take as the high-intensity region all pixels that are above the 95$^{th}$ percentile in integrated line intensity, while our definition of low-intensity gas is all the pixels that are below this same threshold.

Fig.~\ref{fig:density} compares the alignment measure for the low-intensity regions and high-intensity regions ( high-intensity regions for each tracer are indicated with green contours in Fig~.\ref{fig:CO}, Fig.~\ref{fig:HNC}, and Fig.~\ref{fig:CS}), traced by $^{12}$CO, $^{13}$CO, and C$^{18}$O. We find that for VChGs, $^{12}$CO shows little alignment towards the high-intensity  region while C$^{18}$O shows better alignment in high-intensity regions than in low-intensity regions. The weak integrated line intensities of C$^{18}$O towards low column density cloud regions likely contributes to the lower AM values observed towards low column density sight lines.

\subsection{High-density Gas Region}

As is described above, the CS, HNC, HCN, and HCO$^{+}$ are used as dense gas tracers. Although HCN and HCO$^{+}$ are usually generated in non-self-gravitating regions, they can be used to trace gas that is near the self-gravitating regions. Lower density tracers $^{12}$CO $^{13}$CO and intermediate density tracer C$^{18}$O can also be applied to relatively diffuse regions that are near the strong self-gravitating molecular core. 

\begin{figure}
\centering
\includegraphics[width=.99\linewidth,height=0.8\linewidth]{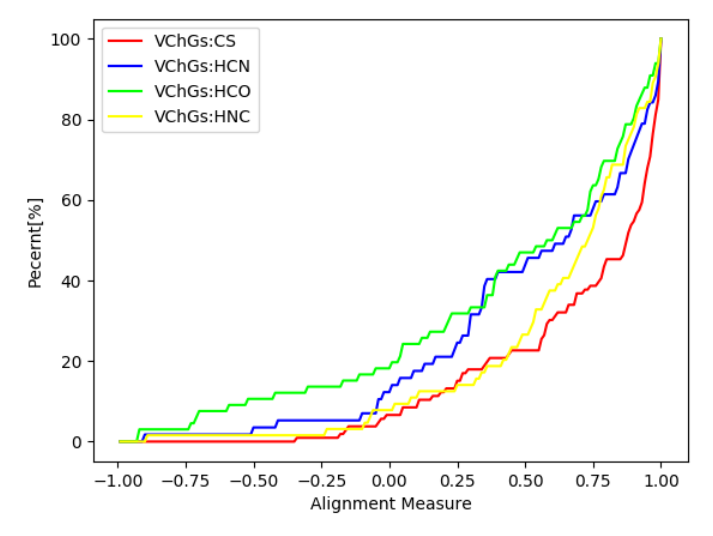}
\caption{\label{fig:variance}The cumulative plot of alignment measure, with respect to different molecular tracers: CS, HCN, HCO$^+$, and HNC. VChGs are used here.}
\end{figure}

From Fig.~\ref{fig:CS} and Fig.~\ref{fig:HNC}, it is obvious that both CS and HNC show excellent alignment between gradients and the magnetic field. Fig.~\ref{fig:variance} shows the cumulative plot of the alignment measure, when using VChGs to trace the magnetic field by CS, HCN, HCO+, and HNC. CS and HNC show a very strong increase in the cumulative fraction over the AM range [0.5,1.0]. 

Therefore, HNC and CS give a more robust performance than others. 
 We therefore propose to use HNC or CS to trace the magnetic field in the dense gas regions (i.e. H$_2$ density $\sim 10^4-10^6 cm^{-3}$) of GMCs.

Furthermore, we propose including HNC to the inferred magnetic field orientation map made from a combination of multiple molecular line tracers. $^{12}$CO, $^{13}$CO, and C$^{18}$O provide the structure of the magnetic field towards low and intermediate column density gas. Taking the dense region traced by HNC which traces higher densities on average than CS into account, we can, therefore, infer the plane of sky component of the magnetic field orientation over a wide range of cloud densities in GMCs. 

\begin{figure}
\centering
\includegraphics[width=0.99\linewidth]{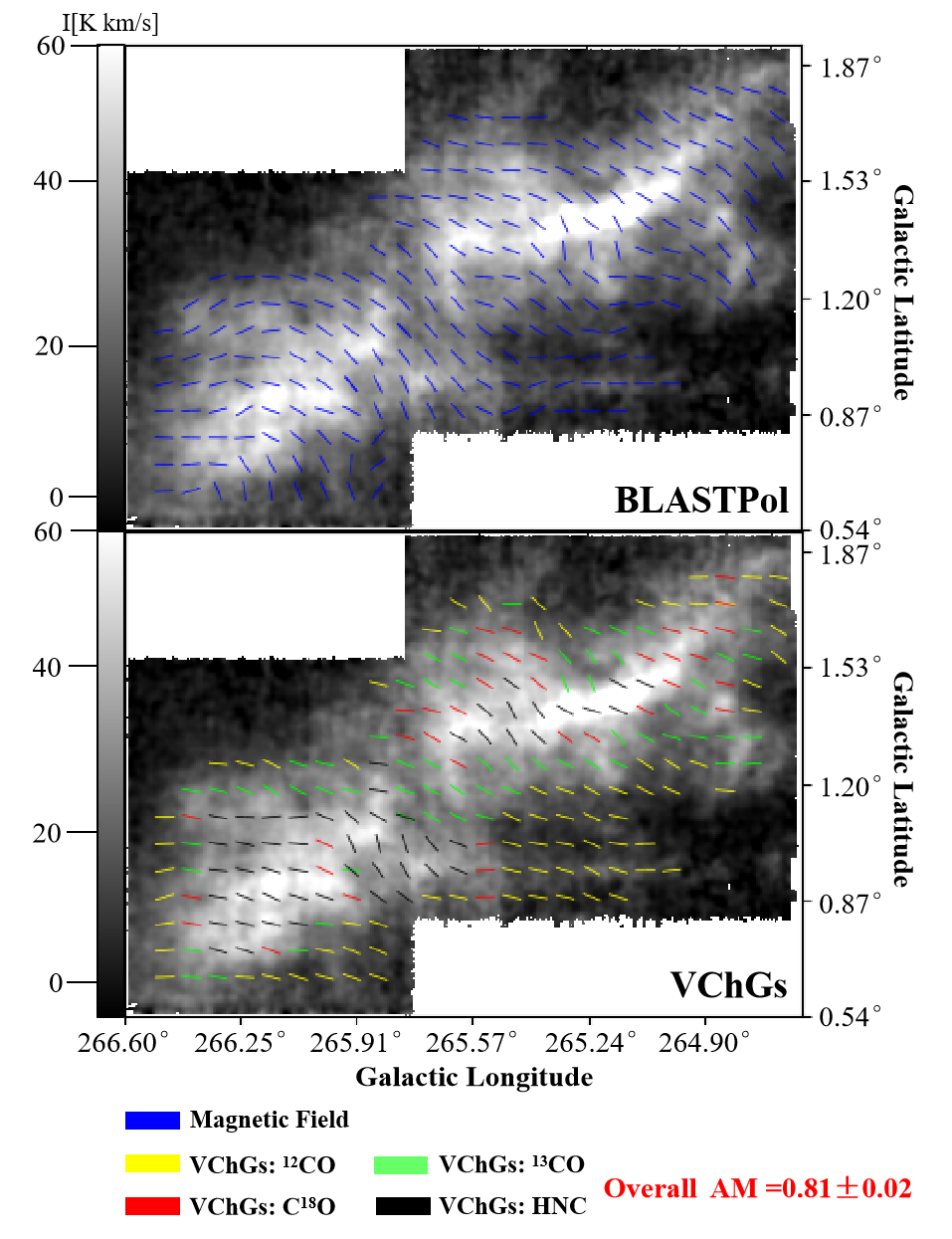}
\caption{\label{fig:3d_all}The top figure is the orientation of magnetic field inferred from BLASTPol 500$\,\mu$m polarization data. The bottom figure shows the planar orientation of magnetic field obtained by rotated VChGs using molecular tracers $^{12}$CO (yellow line segment), $^{13}$CO (lime line segment), C$^{18}$O (red line segment), and HNC (black line segment). The background intensity map is traced by $^{13}$CO.}
\end{figure}
\begin{figure*}
\centering
\includegraphics[width=0.95\linewidth,height=0.75\linewidth]{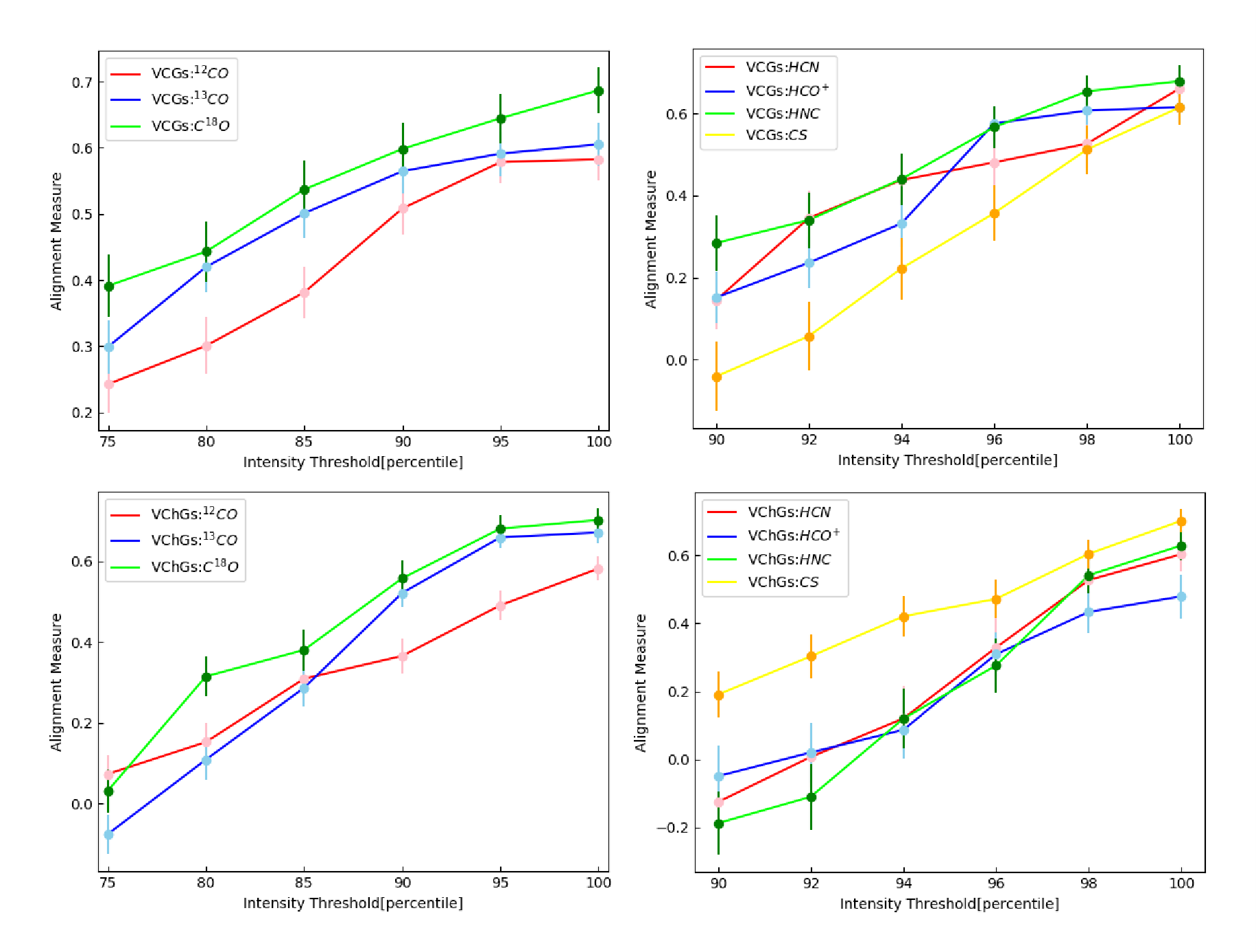}
\caption{\label{fig:threshold}The variation of AM calculated over the whole molecular line map with different re-rotated threshold. Horizontal axis is the threshold for re-rotation. When the density exceeds the threshold, we re-rotate the gradients in that region. The left column is testing with $^{12}$CO, $^{13}$CO, and C$^{18}$O. The right column is CS, HCN, HNC, and HCO$^{+}$. The error bar is given by the standard error of the mean, i.e. the standard deviation divided by the square root of the sample size.}
\end{figure*}
\begin{figure}
\centering
\includegraphics[width=0.99\linewidth]{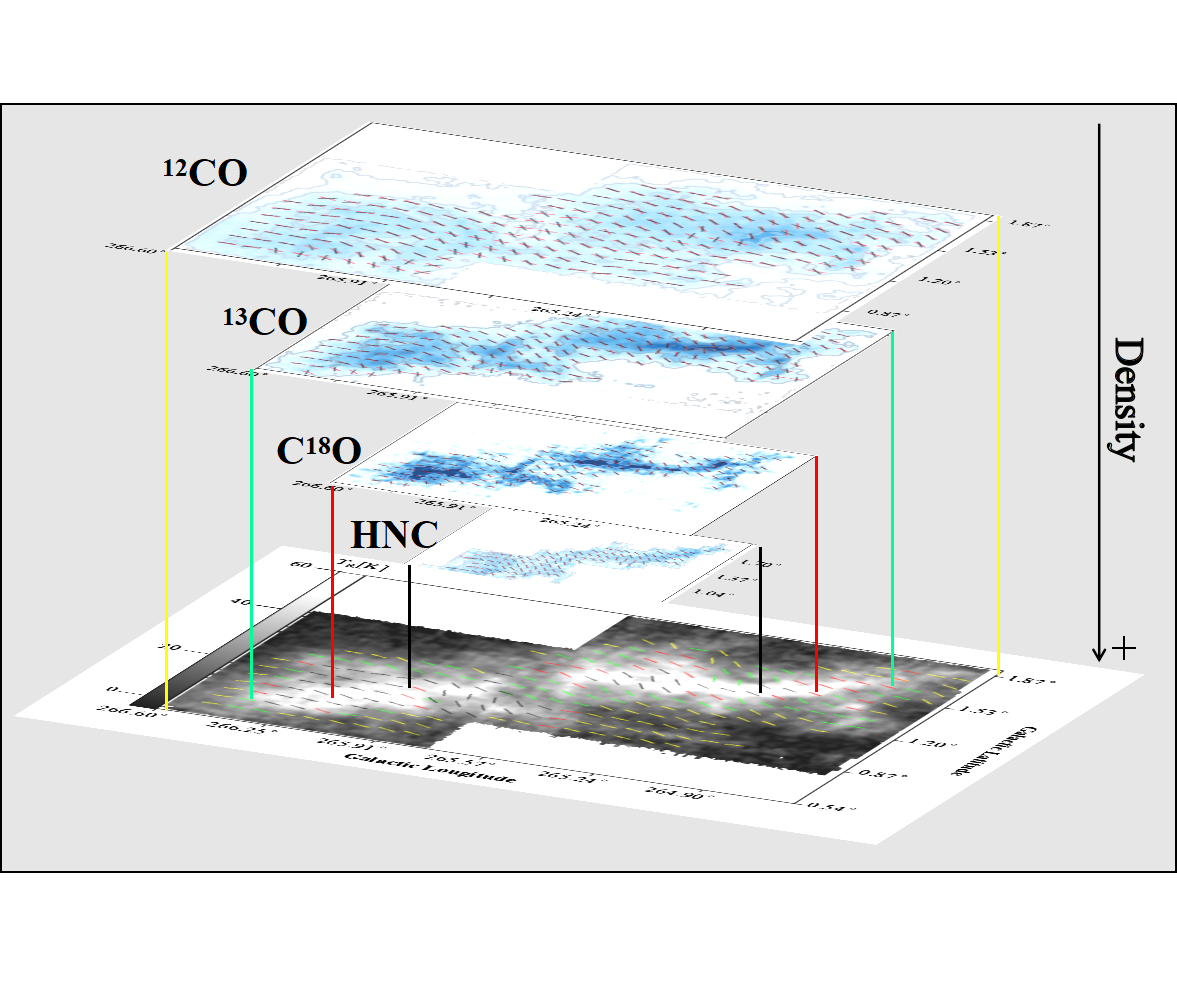}
\caption{\label{fig:3d} The cartoon of gradient tomography maps stacking from $^{12}$CO, $^{13}$CO, C$^{18}$O, and HNC.}
\end{figure}

Fig.~\ref{fig:3d_all} shows the magnetic field structure of Vela C obtained from VChGs using the combination of multi-tracers $^{12}$CO, $^{13}$CO, C$^{18}$O and HNC. The map is produced as follows: using velocity channel maps, we compare the structure contours\footnote{ The structure contour is a curve connecting points which have the same particular value. We use the structure contour to highlight the main structure of Vela C for each tracer, as shown in Fig.~\ref{fig:CO}. The value selected for $^{12}$CO is 50K$\cdot$km/s and above, 15K$\cdot$km/s for $^{13}$CO, 3K$\cdot$km/s for C$^{18}$O, 5K$\cdot$km/s for CS, 4K$\cdot$km/s for HNC, while 2K$\cdot$km/s for HCN and HCO$^+$.} as discussed above from different molecular tracers and remove the gradients calculated from the tracer has lower critical density in the overlap region but keep the gradients calculated from the tracer has higher critical density. The combination of multiple-tracers shows a better alignment measure over all than any single molecular line tracer.

\subsection{Re-rotation Test}
In extremely high density cloud regions, self-gravity can be the main force affecting the dynamics of the gas in GMCs. We expect the gradients to be aligned parallel to the magnetic field with the presence of gravitational collapse, i.e., the direction of the gradient shoud flip 90$^\circ$ \citep{2017arXiv170303026Y}. In this section we search for the signature of gravitational collapse in Vela C using VCGs and VChGs.

We explore this by re-rotating the gradients by 90$^\circ$ again at the high-intensity cloud regions. This is equivalent to not rotating the gradients by 90$^\circ$~towards high intensity regions, but only low intensity regions. For $^{12}$CO, $^{13}$CO, and C$^{18}$O, we vary the threshold for the definition of high-intensity region from the 75$^{th}$ percentile in integrated line intensity,  for which AM is close to 0, to the 100$^{th}$ percentile which means that there is no re-rotation. For CS, HCN, HNC, and HCO$^{+}$, we change the range of high-intensity region from 90$^{th}$ percentile, since with the threshold at the 90$^{th}$ percentile the AM is already close to 0.

Fig.~\ref{fig:threshold} shows the AM variation with different moment = 0 map percentile thresholds defining the re-rotated region. It shows that the peak value of AM among all tracers is achieved at 100$^{th}$ percentile which means no re-rotation is required for both VCGs and VChGs to get good alignment with polarization. Thus the high-intensity region contributes positively to the whole AM value, among all tracers. Therefore we conclude that strongly self-gravitating, collapsing regions constitute only a small fraction of the cloud area within Vela C. This, however, does not prevent molecular gas in small scales regions below the resolution limit of the Mopra data to be collapsing and forming stars \citep{survey}.

\section{Discussion}
\label{sec:discussion}
\subsection{Extracting Magnetic Field Orientation for Gas in Different Density Regimes}
Molecular tracer maps with different optical depth provide the spectroscopic information of gas dynamics up to certain line-of-sight depth. The concept of ”gradient tomography” was first discussed in \citet{2018ApJ...865...59L} by considering the effective accumulation line-of-sight deepness of synchrotron polarization data with the different wavelength. Both the synchrotron polarization data with the presence of strong Faraday Rotation effect and the gas spectroscopic data with the presence of optically thick radiative transfer effect share the same concept that the contribution of gas dynamics with line-of-sight deepness larger than some certain physical boundary would be effectively noise. \citet{2018ApJ...865...59L} showed that, by stacking the gradient maps from the polarized synchrotron intensities measured from different frequencies, one can create the 3D tomography information of the magnetic field. The number of layers in the gradient tomography completely
depends on how many individual frequency measurements one has taken for the synchrotron data. \citet{2019ApJ...873...16H} implemented analogous idea in the case of multiple molecular tracer maps using SPARX radiative transfer code. For our analysis of the Vela C data each molecular tracer samples a different range of densities. Assuming the excitation temperature of Vela C is approximately 10K and CMB brightness temperature 2.725K, \citet{2018arXiv180408979F} then calculated the optical depth $\tau_{18}$ for  C$^{18}$O typically ranges from 0.015 to 0.18, with a median value of 0.026. Assuming a [$^{13}$CO/C$^{18}$O] ratio of 10 and a [$^{12}$CO/ C$^{18}$O] ratio of 400, this implies a typical $\tau_{12}$= [$^{12}$CO/ C$^{18}$O]$\tau_{18}$  in the range of 6 to 72, and $\tau_{13}$ in the range of 0.15 to 1.8. We can therefore expect that the velocity gradients tell us about the plane of sky component of the magnetic field over different density ranges, as shown Fig.~\ref{fig:3d}. 

If the velocity gradient can be used to determine the magnetic field orientation over different cloud density regimes, this also has implications for the efficiency of dust grain alignment as a function of density.  Dust grains in molecular clouds are thought to be aligned by radiative alignment torques (RATs)  (see \citealt{2007MNRAS.378..910L}), and in deeply embedded cloud regions the alignment efficiency may be lower as the photons of the interstellar radiation field with the wavelength comparable with the grain size would be selectively extincted. However, some observational studies and numerical simulation indicate that grains can be effectively aligned for moderately extincted dust sightlines \citep{2007ApJ...663.1055B,2014A&A...569L...1A, 2016ApJ...824..134F}. Moreover, the radiation of the embedded stars can play an important role for the RAT alignment \citep{2008ApJ...674..304W}. 
The fact that the BLASTPol-inferred magnetic field orientation shows a higher degree of alignment with the VGT of intermediate or high density tracers towards high column density sightlines would seem to indicate that dust grains are efficiently aligned in molecular gas with number density $\approx\,10^4\,$cm\,$^{-3}$ or greater, and therefore that dust polarization can trace the magnetic field of intermediate and high density gas. While in numerical simulation it is confirmed that the grain align efficiency is still high fith number density $\ge 10^3$ \citep{2019MNRAS.482.2697S}, this issue requires further studies with higher resolution molecular line observations and polarization data.

\subsection{Applicability of the velocity gradient technique to other molecular clouds}
The gradient technique, which shows very good alignment with the magnetic field orientation obtained from polarization data, is a promising method to infer the magnetic field morphology over a wide range of physical conditions. Low resolution polarization data is now universally available from the Planck all-sky data, while higher resolution polarimetry data has been obtained from stratospheric and ground-based instruments.

However, dust polarization may not be sensitive to the magnetic field within deeply embedded high density regions, where as the efficiency of grain alignment due to radiative torques is expected to be less efficient \citep{2007MNRAS.378..910L}. In addition, it is very difficult to determine the magnetic field structure from polarization data when there are multiple dust clouds along the line of sight.  However, gradient technique can make use of readily available large scale molecular line surveys, such as CHAMP \citep{2006JGRA..111.2304S} and ThRuMMs \citep{2015ApJ...812....7N} survey, or \citep{2016A&A...594A.116H} neutral hydrogen atom distribution survey and the COMPLETE survey \citep{2006AJ....131.2921R}, to provide measurements of the magnetic field orientations, which can also be cross-checked by using different tracers towards the same cloud.

\section{Conclusion }
\label{sec:conclusion}
As one of the most versatile methods for probing magnetic field in the molecular clouds, the gradient method has been proposed to trace magnetic field orientations in multiple scenarios, including diffuse media, shocks, and self-gravitating regions. By examining the gradients of seven chemical tracers, and judging from the distribution of gradient vectors that are rotated by 90$^\circ$, we present here an example making uses of the recent advancement of gradient techniques from numerical and observational studies. Moreover, this work also suggests the promise of using multiple tracers with different tracing abilities to probe magnetic fields in different scales according to the density of the molecular clouds. To summarize:

\begin{enumerate}
\item The Velocity Gradients Technique, which is for the first time applied to multiple tracers, is able to trace the variations in the plane-of-sky component of the magnetic field within different density regimes of the Vela\,C cloud. 

\item The Velocity Gradients Technique opens new a way of exploring the localized magnetic field in Giant Molecular Clouds without the systematic uncertainty inherent in polarization observation associated with subtracting out foreground  and background dust emission.

\item We see no column density threshold above which re-rotation of the velocity gradients is required to match the BLASTPol inferred magnetic field orientation.  This implies that at the resolution of the Mopra data, we do not see the expected signature of the velocity gradients aligning parallel to the magnetic field in regions of gravitational collapse.  We therefore infer that the collapsing regions constitute a small fraction of the Vela C cloud. 
\end{enumerate}

\textbf{Acknowledgements.} A.L. acknowledges the support of the NSF grant AST 1715754, and 1816234 as well as a distinguished visitor PVE/CAPES appointment at the Physics Graduate Program of the Federal University of Rio Grande do Norte, the INCT INEspao and Physics Graduate Program/UFRN. K.H.Y. acknowledges the support of the NSF grant AST 1816234. Y.H. acknowledges the support of the Future Generation Fellowship by Hu's family. L.M.F.~is a Jansky Fellow of the National Radio Astronomy Observatory (NRAO). NRAO is a facility of the National Science Foundation (NSF operated under cooperative agreement by Associated Universities, Inc.). The Mopra radio telescope is part of the Australia Telescope National Facility, which is funded by the Australian Government for operation as a National Facility managed by CSIRO.

\appendix
\section{Statistical error in polarization data}
\label{appendx:A}
\begin{figure}[h]
\centering
\includegraphics[width=0.99\linewidth]{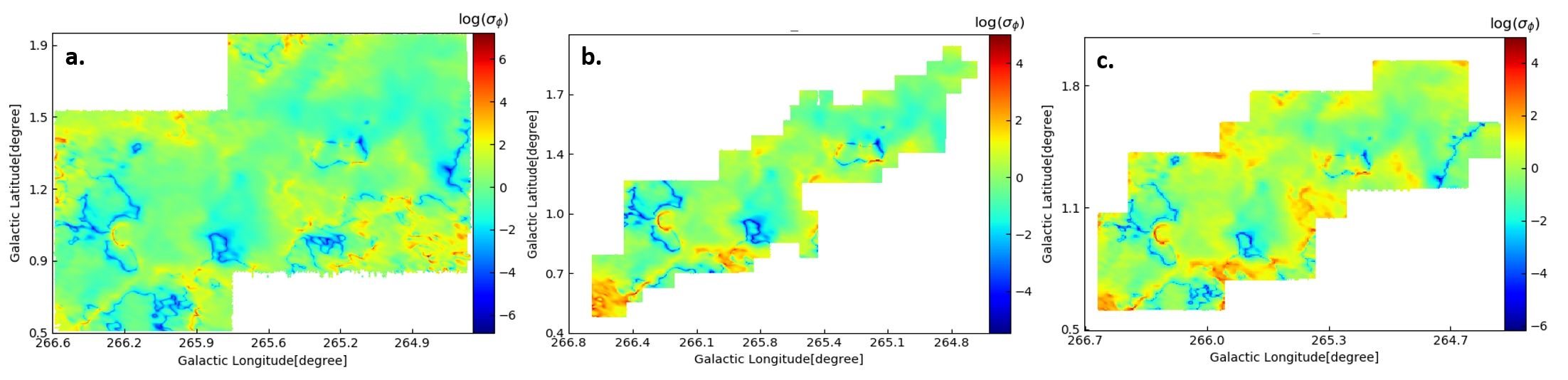}
\caption{\label{fig:var} The statistical angle uncertainty (expressed in degree) in polarization data. \textbf{Panel a}: the polarization data used in the comparison with the CO isotopologues data. \textbf{Panel b}: the polarization data used in the comparison with the HCN, HNC, and HCO$^+$ data. \textbf{Panel c}: the polarization data used in the comparison with the CS data.}
\end{figure}

We show the statistical angle uncertainty in polarization data in Fig.~\ref{fig:var}. The errors are not constant across the BLASTPol map and the error is higher when it gets close to the boundary of Vela C.  However, the mean error for the polarization data used in comparison with the CO isotopologues data is $\sim2.07^\circ$, while $\sim1.44^\circ$ for the one used for CS, and $\sim1.37^\circ$ for the one used for HCN, HNC, and HCO$^+$. The statistical polarization angle errors depend on the signal to noise of the polarization fraction. For Gaussian polarization errors, the error $\sigma_\phi \sim 28.6^\circ \times \sigma_p/p$, where $\phi$ is the polarization angle and p is the polarization fraction \citep{2016ApJ...824..134F}. Thus, for our polarization data, which is a three-sigma polarization detection, the statistical error is less than 10 $^\circ$ \citep{2018arXiv180408979F}.

\end{document}